\documentclass[12pt]{article}
\pdfoutput=1 
\usepackage{graphicx}
\usepackage{hyperref}
\usepackage{lineno}
\usepackage{subcaption}
\usepackage{amsmath}
\usepackage{wrapfig, caption}
\newcommand\pubnumber{ATL-PHYS-PROC-2020-004}
\newcommand\pubdate{\today}
\newcommand\pubblock{\rightline{\begin{tabular}{l} \pubnumber\\ \pubdate \end{tabular}}}
\def\Title   #1{\begin{center}{\Large #1}\end{center}}
\def\Author  #1{\begin{center}{\sc    #1}\end{center}}
\def\Address #1{\begin{center}{\it    #1}\end{center}}
\newenvironment{Abstract} {\begin{quotation}}{\end{quotation}}
\newenvironment{Presented}{\begin{quotation} \begin{center} PRESENTED AT \end{center} \bigskip \begin{center} \begin{large}} {\end{large} \end{center} \end{quotation}}

\def\beq{\begin{equation}}
\def\eeq#1{\label{#1}\end{equation}}
\def\eeqn{\end{equation}}

\def\beqa{\begin{eqnarray}}
\def\eeqa#1{\label{#1}\end{eqnarray}}
\def\eeqan{\end{eqnarray}}

\let\bar=\overbar

\def\Dslash{\not{\hbox{\kern-4pt $D$}}}
\def\dslash{\not{\hbox{\kern-2pt $\del$}}}

\def\msb{{\bar{\ssstyle M \kern -1pt S}}}

\begin{document}
\begin{titlepage}
  
  \pubblock

  \vfill \Title{Overview of Recent Results from ATLAS} \vfill
  \Author{Craig Wiglesworth, on behalf of the ATLAS Collaboration}
  \Address{Niels Bohr Institute, Blegdamsvej 17, 2100 Copenhagen, Denmark}
  
  \vfill \begin{Abstract} This paper presents an overview of recent results from the ATLAS Experiment at the CERN Large Hadron Collider, with a particular focus on those that are based on the entire Run 2 dataset of $\sqrt{s}$=13~TeV proton-proton collisions. These results include Standard Model measurements, updates on the Higgs Boson properties, as well as searches for high-mass resonances and supersymmetry.
 \end{Abstract}

  \vfill \begin{Presented} $12^\mathrm{th}$ International Workshop on Top Quark Physics\vspace{2mm}\\ Beijing, China, September 22--27, 2019 \end{Presented} \vfill
  
\end{titlepage}
\section*{Introduction}

The Large Hadron Collider (LHC) \cite{Evans:2008zzb} has performed exceptionally well in 2018, achieving a peak instantaneous luminosity of 2.1x10$^{34}$~cm$^{-2}$s$^{-1}$ \cite{TWIK-LUMI-000} - more than twice its design luminosity. Similarly, the ATLAS Detector \cite{Aad:2008zzm} continues to improve year-on-year, with data-taking and data-quality efficiencies in 2018 of 95.7\% \cite{TWIK-LUMI-000} and 97.5\% \cite{TWIK-DATQ-000}, respectively. Across the entire Run-2 data-taking period (2015-2018) ATLAS has collected a dataset of $\sqrt{s}=$13~TeV proton-proton collisions, corresponding to an integrated luminosity of 139~fb$^{-1}$ \cite{TWIK-LUMI-000}.
\\
\\
The ATLAS Experiment has a broad physics programme that includes: probing the Standard Model (SM), precision measurements of the Higgs boson and searches for physics beyond the SM (BSM), such as supersymmetry (SUSY) and other exotic phenomena. This paper presents a non-comprehensive selection of recent results which, unless otherwise stated, are based on the entire Run~2 dataset of $\sqrt{s}$=13~TeV proton-proton collisions. In addition to the statistical improvements owing to a larger dataset, a number of systematic improvements in reconstruction and analysis techniques have contributed to increased sensitivities with respect to previous ATLAS results.

\section*{Overview of Recent Results}
Following the discovery of the Higgs boson, scrutiny of the electroweak symmetry breaking (EWSB) has become a major focus at the LHC. In addition to direct measurements of Higgs boson properties, the study of massive vector-boson scattering (VBS) offers another key avenue to probe EWSB. The signature of VBS is a pair of vector-bosons and two hadronic jets, denoted as $VVjj$. The electroweak production of $WWjj$ and $WZjj$ have already been observed in ATLAS \cite{Aaboud:2019nmv,Aaboud:2018ddq}, with no obvious deviations from the predictions of the SM. A complementary search for the electroweak production of $ZZjj$ has been performed, in which the 4-lepton and 2-lepton 2-neutrino final states were considered \cite{ATLAS:2019vrv}. The search uses multivariate discriminants based on event kinematics, trained separately in each channel to enhance the separation between the signal and backgrounds. By combining the two search channels a deviation from the background-only hypothesis is observed, corresponding to a statistical significance of 5.5 standard deviations ($\sigma$). This observed excess is consistent with the electroweak production of a $Z$ boson pair in association with two hadronic jets and provides the first observation of the electroweak production of $ZZjj$.
\\
\\
The large cross section for $t\bar{t}$ production at the LHC allows ATLAS to measure the properties of the top-quark through the study of its decay products. One such property is the $t\bar{t}$ charge asymmetry $A_{C}$, representing the central-forward charge asymmetry that originates from the interference of higher-order amplitudes in $q\bar{q}$ and $qg$ initial states. $A_{C}$ is defined as,

\begin{equation*}
  A_{C} = \frac{N(\Delta|y| > 0) - N(\Delta|y| < 0)}{N(\Delta|y| > 0) + N(\Delta|y| < 0)}
\end{equation*}
\vspace{0.5mm}
\begin{wrapfigure}{r}{0.5\textwidth}
  \centering
  \captionsetup{width=0.45\textwidth}
  \includegraphics[width=0.5\textwidth]{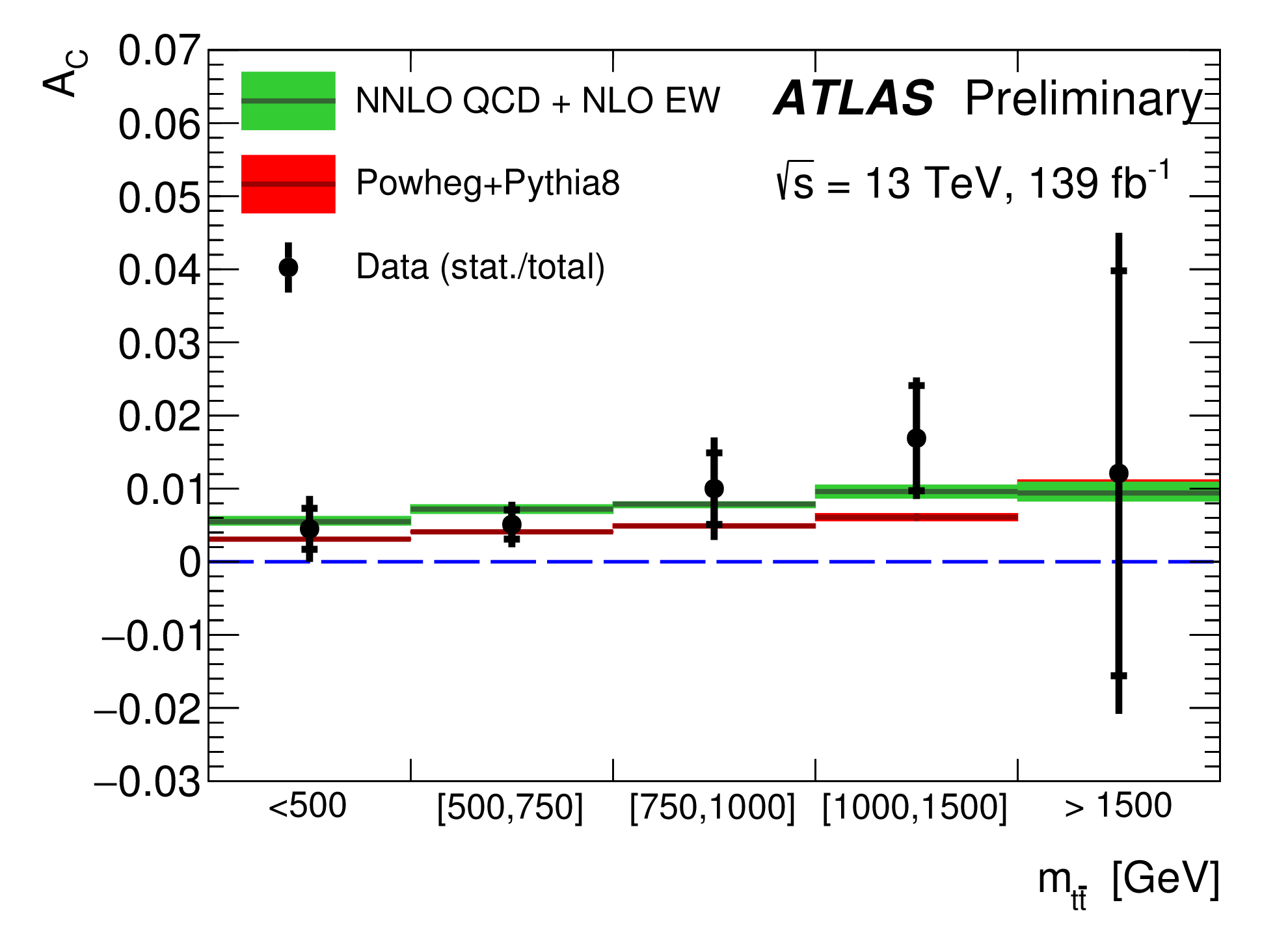}
  \caption{The measured charge asymmetry as a function of the invariant mass of the $t\bar{t}$ system. The green regions show the SM predictions, whilst the red regions show the charge asymmetry in simulated events \cite{ATLAS:2019czt}.}
  \label{fig:topq}
\end{wrapfigure}
where $\Delta|y| = |y_{t}|-|y_{\bar{t}}|$ is the difference in the absolute values of rapidity for $t$ and $\bar{t}$. Several BSM models predict charge asymmetries that vary as a function of the invariant mass ($m_{t\bar{t}}$) and the longitudinal boost along the z-axis ($\beta_{z,t\bar{t}}$) of the $t\bar{t}$ system. A measurement of $A_{C}$ has been performed in the single lepton channel, combining both the resolved and boosted topologies of top quark decays \cite{ATLAS:2019czt}. A Bayesian unfolding procedure is used to infer the asymmetry at parton level, correcting for detector resolution and acceptance effects. The inclusive $t\bar{t}$ charge asymmetry is measured to be $A_{C}$~=~0.0060~$\pm$~0.0015, which differs from zero by 4~$\sigma$ and provides the first evidence of $t\bar{t}$ charge asymmetry at the LHC. Differential measurements have also been performed as a function of $m_{t\bar{t}}$ and $\beta_{z,t\bar{t}}$. Both the inclusive and differential measurements are found to be compatible with the predictions of the SM, as illustrated in Figure~\ref{fig:topq} which shows the differential measurement as a function of $m_{t\bar{t}}$.

\begin{figure}[h]
  \begin{subfigure}{0.6\textwidth}
    \raggedright
    \includegraphics[width=0.90\textwidth]{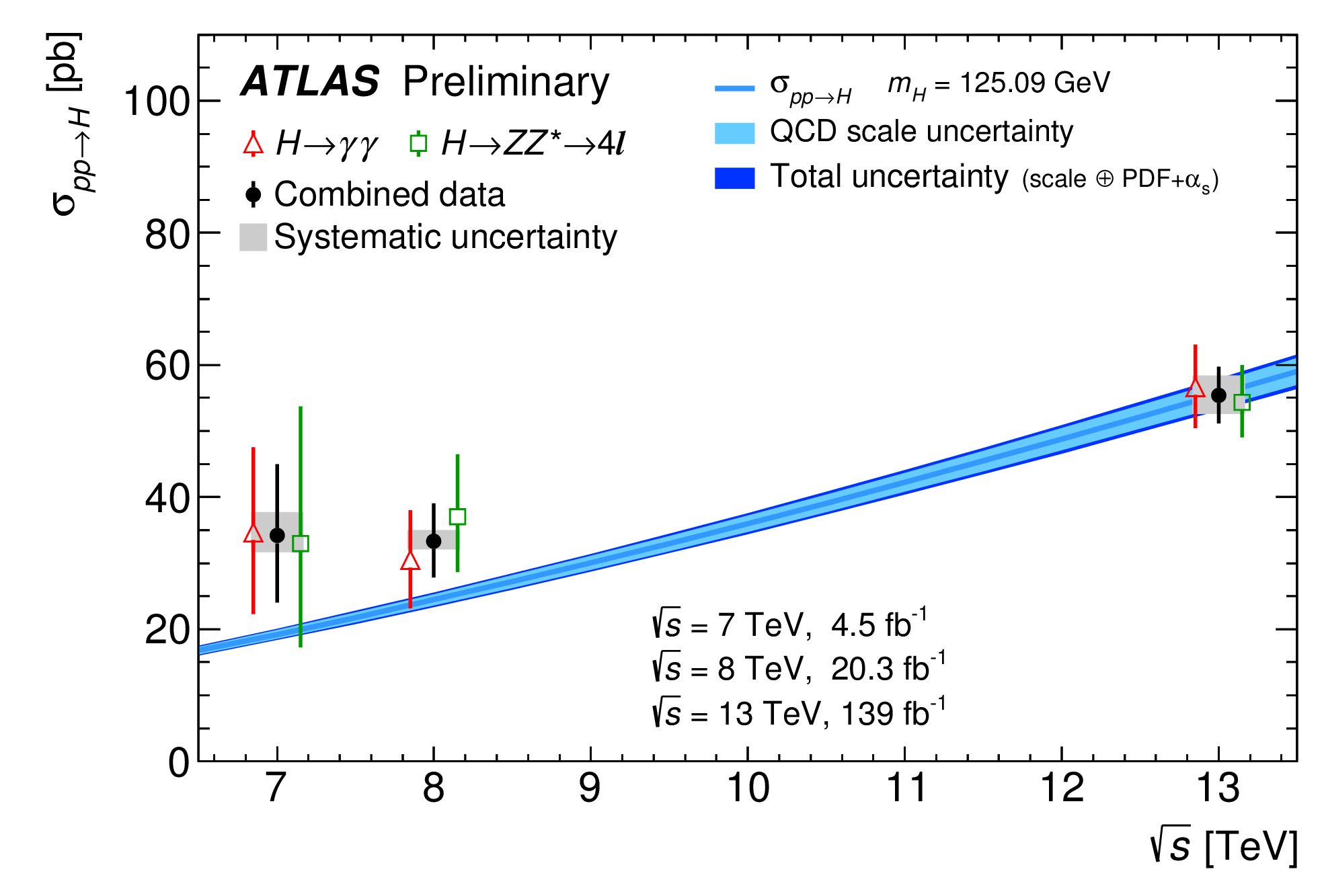}
  \end{subfigure}
  \begin{subfigure}{0.4\textwidth}
    \raggedleft
    \vspace{-3mm}
    \includegraphics[width=0.91\textwidth]{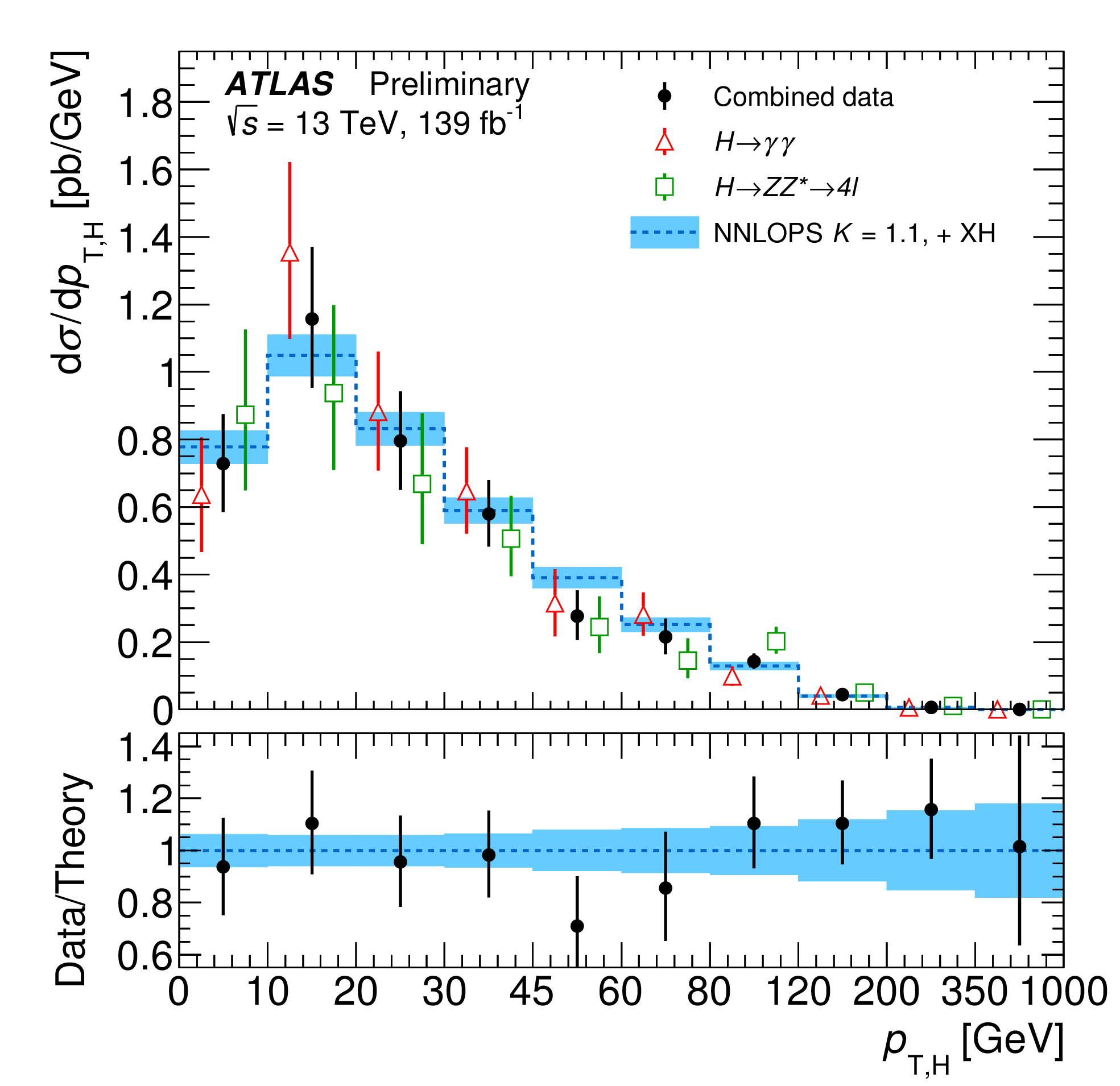}
  \end{subfigure}
  \caption{The measured total production cross section along with those measured at $\sqrt{s}$=7 and 8~TeV (left) and the differential production cross section as a function of the Higgs boson transverse momentum $p_{T,H}$ (right). In both cases the predictions of the SM are shown for comparison \cite{ATLAS:2019mju}.}
  \label{fig:higg}
\end{figure}

A direct measurement of the top-quark decay width ($\Gamma_{t}$) has also been performed in dileptonic decays of $t\bar{t}$ events \cite{ATLAS:2019onj}. Many BSM models predict a top-quark decay width that differs from the prediction of the SM. Any deviation in the top-quark decay width from the expected SM value could therefore hint at new physics, such as non-SM decays of the top-quark or modifications of top-quark couplings. The measurement is based on the invariant mass of the charged lepton and the corresponding $b$-jet, which is used in a profile-likelihood fit with simulated templates for various top-quark decay widths. The decay width is measured to be $\Gamma_{t}$~=~1.9~$\pm$~0.5~GeV, in agreement with the prediction of the SM.
\\
\\
All major production and decay modes of the Higgs boson have now been observed in ATLAS, as we move into the realm of precision Higgs physics. Combined measurements of the total and differential Higgs boson production cross sections have been performed \cite{ATLAS:2019mju}, based on recently updated measurements in the $H{\rightarrow}\gamma\gamma$ \cite{ATLAS:2019jst} and $H{\rightarrow}ZZ^{*}{\rightarrow}$4$l$ \cite{ATLAS:2019ssu} decay modes. The total Higgs boson production cross section is measured to be 55.4$^{+4.3}_{-4.2}$~pb, corresponding to an uncertainty of approximately 8\%. The cross sections measured in each of the individual decay modes and the combined measurements are all in agreement with the SM prediction, as shown in Figure~\ref{fig:higg}.

The interactions between the Higgs boson and charged, third generation fermions have already been observed at the LHC although there is, as of yet, no evidence of Higgs boson interactions with first or second generation fermions. A search for the $H{\rightarrow}\mu\mu$ decay has been performed \cite{ATLAS:2019ain}, in which any deviation from the prediction of the SM could be a sign of new physics. No significant excess is observed above the expected background and the observed upper limit at the 95\% Confidence Level (CL) on the cross section times branching ratio is 1.7 times the SM prediction. Similar searches have been performed for the $H{\rightarrow}ee$ decay and the lepton-flavour-violating $H{\rightarrow}e\mu$ decay \cite{Aad:2019ojw}. In both cases no significant excess is observed above the expected background - in agreement with the SM expectation. The observed upper limits at the 95\% CL on the branching fractions of $H{\rightarrow}ee$ and $H{\rightarrow}e\mu$ are 3.6x10$^{-4}$ and 6.1x10$^{-5}$, respectively. These results represent considerable improvements in sensitivity with respect to previous measurements. 

The production of a pair of Higgs bosons ($HH$) is a rare process in the SM, although a number of BSM models predict cross sections for $HH$ production that are much larger than that of the SM. A search for the $HH{\rightarrow}b\bar{b}b\bar{b}$ process via Vector Boson Fusion (VBF) production has been performed using 126~fb$^{-1}$ of $\sqrt{s}=$13~TeV proton-proton collision data \cite{ATLAS:2019dgh}. The search is sensitive to VBF production of additional heavy bosons that may decay to Higgs boson pairs and, in a non-resonant topology, can constrain the 4-point ($VVHH$) coupling, $c_{2V}$. No significant excess is observed above the expected background. Upper limits are set on the cross section for resonant $HH$ production in the range 260~GeV to 1000~GeV at the 95\% CL. The observed upper limit at the 95\% CL on the cross section for non-resonant production is 1600~fb, with the observed excluded region corresponding to $c_{2V}<$-1.02 and $c_{2V}>$2.71.
\\
\\
A number of exotic BSM models such as Grand Unified models predict the existence of heavy particles, whose decays could be observed as a resonance or a localised excess on top of a smoothly falling background distribution. ATLAS has performed a number of searches for such heavy particles but, as of yet, no excess has been observed above the expected background.

A search for high-mass dielectron and dimuon resonances has been performed and, in the context of various benchmark models, lower limits are set on the mass of a $Z'$ resonance \cite{Aad:2019fac}. In particular, combined lower limits of 4.5~TeV and 5.1~TeV are set at 95\% CL on the $Z'_{\psi}$ of an E$_{6}$-motivated Grand Unified model and the $Z'_{SSM}$ of the Sequential SM (SSM), respectively. These are the most stringent limits to date.

A search for a heavy charged boson resonance decaying to a charged lepton and a neutrino has also been performed \cite{Aad:2019wvl}. Lower limits of 6.0~TeV and 5.1~TeV on the $W'$ boson mass are set at 95\% CL in the electron and muon channels, respectively, in the context of the SSM. To allow for further interpretations of the results, a set of model-independent upper limits are determined for the number of signal events and for the visible cross section above a given transverse mass threshold.

Excited quarks ${q^{*}}$ are predicted in models of compositeness and are a typical benchmark for quark-gluon resonances. A search for high-mass resonances decaying into two hadronic jets has been performed \cite{Aad:2019hjw}. Excited quarks with masses below 6.7 TeV are excluded at the 95\% CL. The results substantially extend the excluded ranges by approximately 700 GeV, with respect to the mass limits of ${q^{*}}$ models obtained in previous searches.

Heavy vector boson resonances are predicted in several extensions to the SM. A search for such diboson resonances in the fully hadronic final state has been performed \cite{Aad:2019fbh} and exclusion limits are set on resonances in a range of BSM models. The highest excluded mass at 95\% CL for a heavy vector boson resonance is 3.8~TeV, in the context of a benchmark heavy vector triplet model in which resonances couple predominantly to gauge bosons.
\\
\\
Extensions of the SM involving SUSY are appealing from a theoretical perspective, due to their potential to alleviate the problem of naturalness. A number of searches have been performed for both strong and electroweak SUSY in ATLAS but, as of yet, no significant excess has been observed above the expected background.

\begin{figure}[h]
  \begin{minipage}{0.5\textwidth}
    \captionsetup{width=0.9\textwidth}
    \raggedright
    \includegraphics[width=0.97\textwidth]{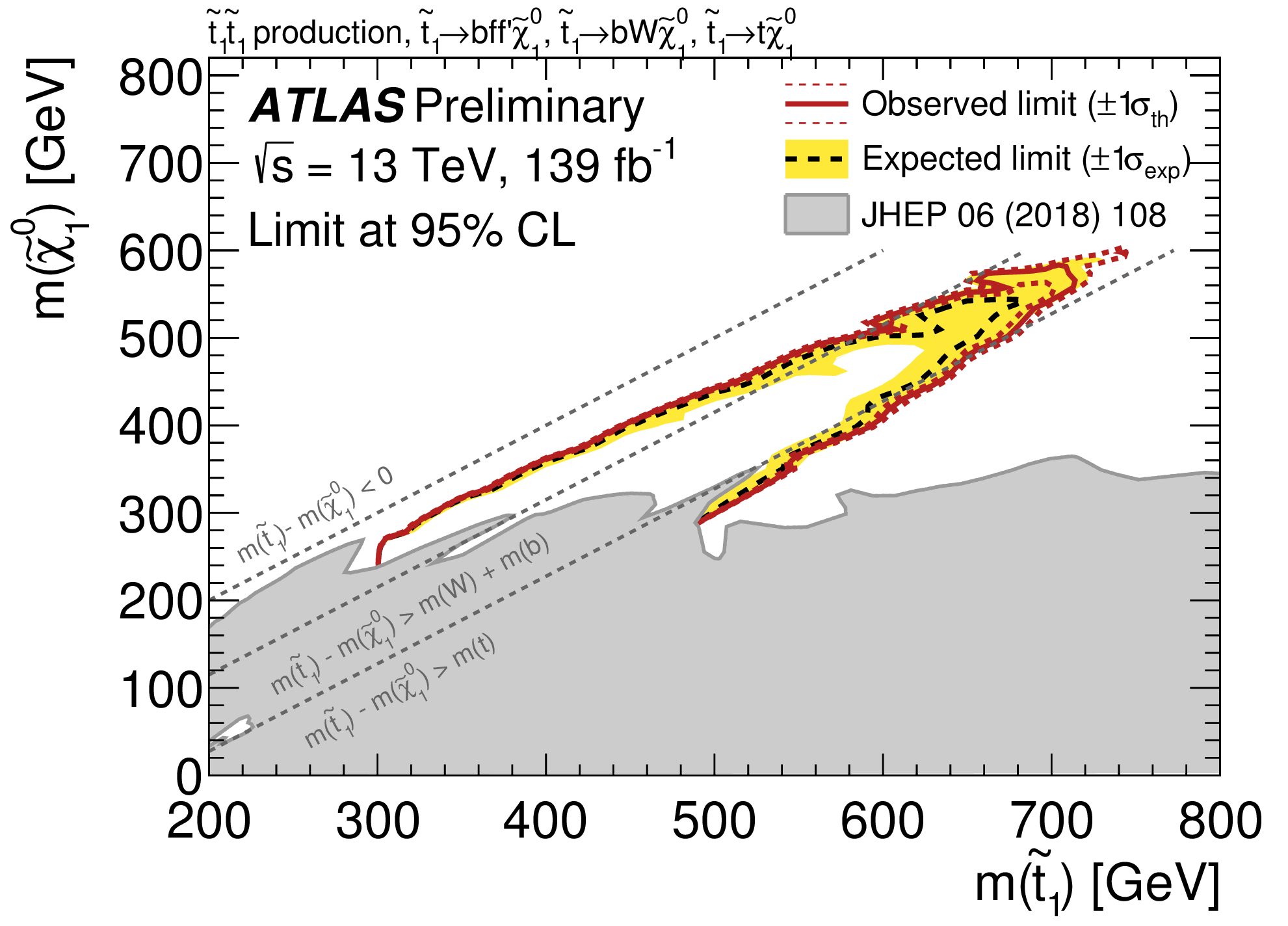}
    \caption{Exclusion limits on top-squark pair production. The mass of the top-squark is shown on the x-axis, while the mass of the neutralino is shown on the y-axis \cite{ATLAS:2019oho}.}
    \label{fig:susy_a}
  \end{minipage}
  \begin{minipage}{0.5\textwidth}
    \captionsetup{width=0.85\textwidth}
    \vspace{-0.5mm}
    \raggedleft
    \includegraphics[width=0.92\textwidth]{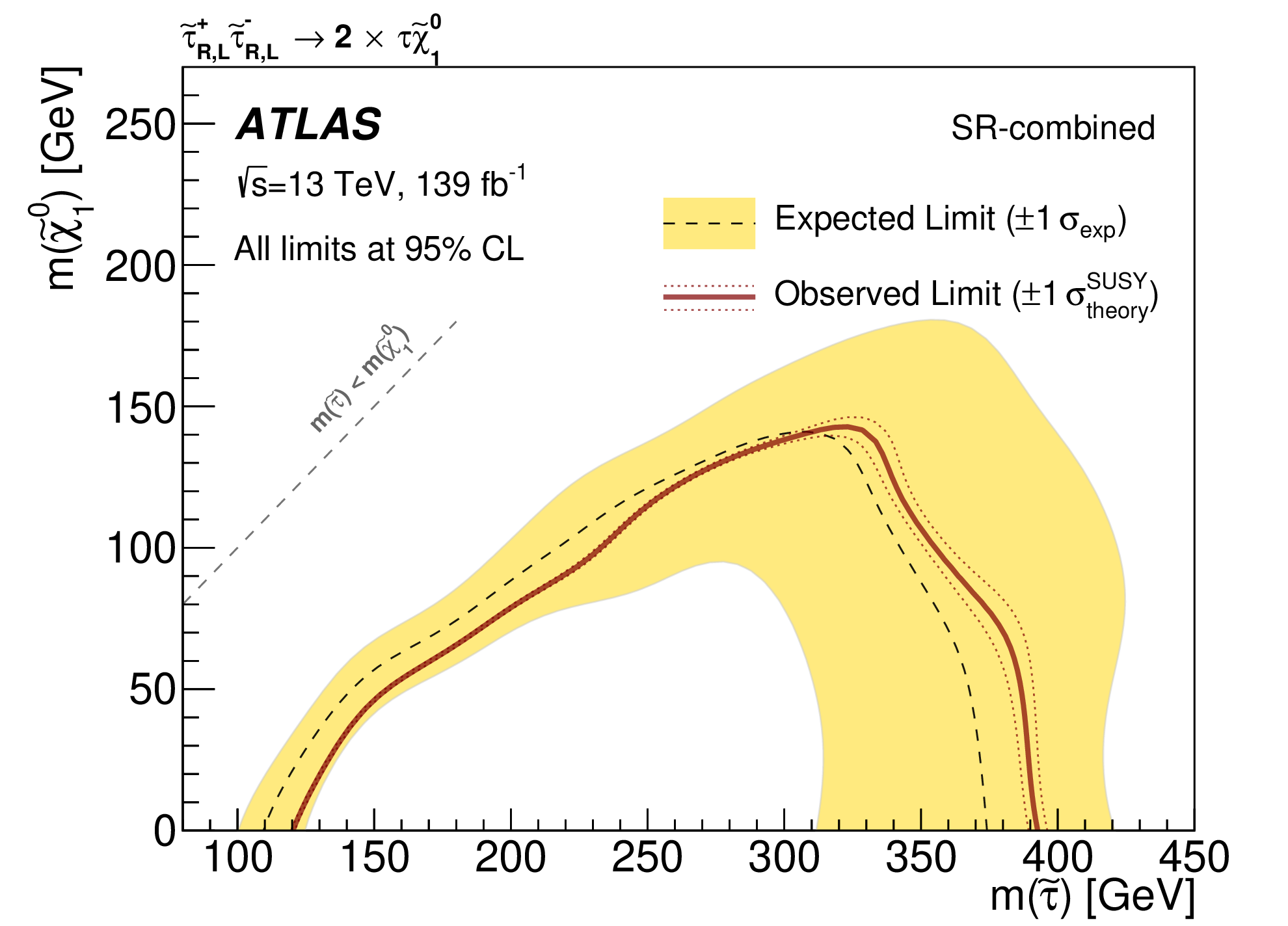}
    \vspace{3.2mm}
    \caption{Exclusion limits on tau-slepton pair production. The mass of tau-slepton is shown on the x-axis, while the mass of the neutralino is shown on the y-axis \cite{Aad:2019byo}.}
    \label{fig:susy_b}
  \end{minipage}
\end{figure}

The supersymmetric partner of the top-quark, the top-squark, is of particular interest as it may largely cancel divergent loop corrections to the Higgs-boson mass. A search for the direct production of top-squark pairs has been performed, focusing on one particular scenario whereby each top-squark decays via a 3-body process to a $b$-quark, a $W$ boson and a neutralino \cite{ATLAS:2019oho}. Exclusion limits at the 95\% CL are set on the considered benchmark model, as shown in Figure~\ref{fig:susy_a}. The result excludes top-squark masses up to 720~GeV with neutralino masses up to 580~GeV.

SUSY models with a light tau-slepton, the supersymmetric partner of the tau lepton, are also of interest since they can lead to a dark matter relic density consistent with cosmological observations. A search for the direct production of tau-slepton pairs in final states with two hadronically decaying tau leptons has been performed \cite{Aad:2019byo}. Exclusion limits are set at the 95\% CL on a simplified model in which two mass-degenerate tau-sleptons each decay into a tau lepton and a neutralino, as shown in Figure~\ref{fig:susy_b}. The result extends the lower mass limit of 86.6~GeV previously set at the Large Electron Positron collider (LEP).

Many other BSM models predict long-lived, weakly interacting particles that can generate a variety of unconventional signatures for which there are often no irreducible SM backgrounds. A search for long-lived particle decays involving a displaced vertex and displaced muon has been performed using 136~fb$^{-1}$ of $\sqrt{s}=$13~TeV proton-proton collision data \cite{ATLAS:2019ems}. Exclusion limits are set in the context of a SUSY model in which a pair of long-lived top-squarks decay via a small R-parity-violating coupling to a quark and a muon. Top-squarks with masses up to 1.7~TeV for a lifetime of 0.1~ns and below 1.3~TeV for all lifetimes between 0.01~ns and 30~ns, are excluded at the 95\% CL.
\\
\\
In addition to the many results based on proton-proton collision data, a search for the rare process of light-by-light scattering has been performed in a data sample of $Pb$+$Pb$ collisions at $\sqrt{s_{NN}}$ = 5.02~TeV \cite{Aad:2019ock}. Light-by-light scattering can occur in relativistic heavy-ion collisions at impact parameters larger than approximately twice the radius of the ions where, in these ultra-peripheral collision events, the strong interaction becomes less significant than the electromagnetic interaction. In the SM the process occurs at one-loop level at order $\alpha_{em}^{4}$, via virtual box diagrams involving electrically charged fermions or $W^{\pm}$. However, in various extensions of the SM, extra contributions are possible, making the process sensitive to new physics. Light-by-light scattering events are characterised by the exclusive production of two low-energy photons in an otherwise empty event. In a dataset of 1.73~nb$^{-1}$ an excess of such events is observed with respect to the background-only hypothesis. The observed excess corresponds to a statistical significance of 8.2~$\sigma$ and is consistent with the SM prediction. This result demonstrates the power and flexibility of the ATLAS Detector, which was primarily designed for very different event topologies.

\section*{Conclusions}

An overview of recent results from the ATLAS Experiment at the CERN Large Hadron Collider has been presented, including some of the first results based on the entire Run 2 dataset of $\sqrt{s}$~=~13~TeV proton-proton collisions. Highlights include: the observation of the electroweak production of a $Z$ boson pair in association with two hadronic jets; the first evidence of charge asymmetry in $t\bar{t}$ pairs at the LHC; precision measurements of the Higgs boson production cross section and the observation of light-by-light scattering in $Pb$+$Pb$ collisions. So far, all experimental results have been consistent with the predictions of the Standard Model.

\\
\\
\noindent Copyright CERN for the benefit of the ATLAS Collaboration. CC-BY-4.0 license. 


\end{document}